\documentclass[prb,twocolumn,floatfix,twoside,tbtags]{revtex4}

\usepackage[latin1]{inputenc}
\usepackage{subfigure}
\usepackage{graphicx}
\usepackage{amsmath,amssymb}

\newcommand{\ie}{\textit{i.e.}{ }}

\DeclareMathOperator{\q}{q}
\newcommand{\me}{\mathrm{e}}
\newcommand{\xeq}{x_{\mathrm{X}}^{\mathrm{eq}}}
\newcommand{\xO}{x_{\mathrm{X}}^{0}}
\newcommand{\oun}{\omega^{(1)}_{\mathrm{AlX}}}
\newcommand{\odeux}{\omega^{(2)}_{\mathrm{AlX}}}
\newcommand{\dmu}{\delta\mu}

\begin{document}

\title{Classical nucleation theory in ordering alloys precipitating with L1$_2$ structure.}

\author{Emmanuel \surname{Clouet}}
\email{emmanuel.clouet@cea.fr}
\affiliation{Service de Recherches de Métallurgie Physique, CEA/Saclay,
91191 Gif-sur-Yvette, France}

\author{Maylise \surname{Nastar}}
\affiliation{Service de Recherches de Métallurgie Physique, CEA/Saclay,
91191 Gif-sur-Yvette, France}


\date{\today}
\begin{abstract}
	By means of low-temperature expansions (LTEs),
	the nucleation free energy and the precipitate interface free energy
	are expressed as functions of the solubility limit
	for alloys which lead to the precipitation
	of a stoichiometric  L1$_2$ compound 
	such as Al-Sc or Al-Zr alloys.
	Classical nucleation theory is then used to obtain
	a simple expression of the nucleation rate whose validity is demonstrated
	by a comparison with atomic simulations.
	LTEs also explain why simple mean-field approximation
	like the Bragg-Williams approximation fails to predict correct nucleation rates 
	in such an ordering alloy.
\end{abstract}
\maketitle

Since its initial formulation in 1927 by Volmer, Weber and Farkas 
and its modification in 1935 by Becker and Döring 
the classical nucleation theory (CNT) \cite{MAR78,KEL91,KAS00} has been a suitable tool 
to model the nucleation stage in phase transformations. 
The success of this theory relies on its simplicity 
and on the few parameters required to predict the nucleation rate. 
Recently, the use of computer simulations 
have allowed to assess the applicability
of the theory for solid-solid phase transformations 
\cite{SHN99,SOI00,CLO04,BER04}. 
Thanks to a precise control of simulation conditions, 
it is possible to get accurate estimations of CNT parameters 
and thus to make a direct comparison 
between theory predictions and quantities observed during simulations.
One thus gains a deeper understanding of the validity of the different
assumptions used by the CNT.

Previous studies have shown that the capillary approximation, which CNT relies on, 
gives a precise description of cluster thermodynamics.
Within this approximation, the free energy of a nucleus 
is written as the sum of a volume contribution,
the nucleation free energy, and a surface contribution 
corresponding to the energy cost to create an interface between 
the nucleus and the solvent. 
For CNT to agree with atomic simulations, 
care has to be taken in the way these two energetic contributions are obtained.
In particular, we have shown that one has to take into account 
short range order when calculating the nucleation free energy 
in an ordering alloy \cite{CLO04}.
Usual thermodynamic approximations, like the ideal solid solution or the Bragg-Williams 
approximation, cannot describe short range order and thus can predict 
values of the cluster size distribution and of the nucleation rate 
wrong by several orders of magnitude.
This is to contrast with more sophisticated mean-field approximations like the cluster 
variation method (CVM) which provides good predictions of the nucleation rate \cite{CLO04}.
However, an easy use of CNT and a clear determination of the missing ingredients 
in simple mean-field approximations  requires an analytical approach 
which CVM cannot provide.
Such an approach has to lead to accurate expressions of the CNT input parameters
so as to make the theory predictive without any fitting of its parameters.

In this Letter, we use low-temperature expansions (LTE) \cite{DUCASTELLE,DOMB3}
to derive an analytical formulation of the nucleation free energy and the interface free energy
in a binary system like Al-Sc or Al-Zr, 
\ie a supersaturated Al-X solid solution leading
to the nucleation of a stoichiometric Al$_3$X compound with the L1$_2$ structure.
This structure corresponds to an ordering of the fcc lattice with solute X atoms
lying on one of the four cubic sublattices\cite{DUCASTELLE}.
LTE are well suited to describe short range order in dilute solid solution 
and nearly stoichiometric ordered compounds \cite{KOH98,AST98,HYL98,WOO01,LEB03}
like Al$_3$X compound.
The use of this method in CNT framework allows to obtain 
a fully analytical modelling whose only material parameters are the solubility limit 
and the solute diffusion coefficient.

To do so, we start from the same atomic diffusion model previously
developed for Al-Sc-Zr system \cite{CLO04,CLO06}. 
This model relies on a rigid lattice with interactions between 
first- and second nearest neighbors and uses a thermally activated
atom-vacancy exchange mechanism to describe diffusion. 
Despite its simplicity, it has been shown to lead to predictions 
in good agreement with experimental data \cite{CLO06,CLO05,CLO07}.
Within this atomic model, atoms are constrained to lie on a fcc lattice
and the configurations of a binary Al-X alloy is fully described by
the solute atom occupation number $p_n$ with $p_n=1$ if the site $n$
is occupied by a solute atom and $p_n=0$ otherwise.
The energy per atom of a given configuration of the Al-X alloys is then given by
\begin{multline}
	E = U_{\mathrm{Al}} 
	+ \frac{1}{N_s} \left( U_{\mathrm{X}}-U_{\mathrm{Al}} \right)\sideset{}{_n}\sum{p_n} \\
	+ \frac{1}{N_s} \oun   \sideset{}{_{n,m}^{'}}\sum{\left( 1-p_n \right)p_m} \\
	+ \frac{1}{N_s} \odeux \sideset{}{_{n,m}^{''}}\sum{\left( 1-p_n \right)p_m} \label{eq:Ising}
\end{multline}
where the first and second sums, respectively, run on all first and second nearest-neighbor pairs of sites,
$N_s$ is the number of lattice sites,
$U_{\mathrm{Al}}$ (respectively $U_{\mathrm{X}}$) is the energy per atom
when only Al (respectively X) atoms lie on the fcc lattice and
$\oun$ and $\odeux$ are the first and second nearest neighbor order energies.
Al-Sc and Al-Zr thermodynamics are characterized by the order tendency 
between first nearest neighbors ($\oun<0$) 
and the demixing tendency between second nearest neighbors ($\odeux>0$).
Eq.~\ref{eq:Ising} is a rewriting for binary alloys
of the atomic model developed in Refs.~\cite{CLO04,CLO06}
when one neglects vacancy contributions.

The nucleation free energy entering CNT is defined by 
\begin{multline}
	\label{eq:Gnuc}
	\Delta G^{\mathrm{nuc}}(x^0_{\mathrm{X}}) = 
	\frac{3}{4} \left[ \mu_{\mathrm{Al}}(x^{eq}_{\mathrm{X}})
	- \mu_{\mathrm{Al}}(x^{0}_{\mathrm{X}}) \right] \\
	 + \frac{1}{4} \left[ \mu_{\mathrm{X}}(x^{eq}_{\mathrm{X}})
	- \mu_{\mathrm{X}}(x^{0}_{\mathrm{X}}) \right] ,
\end{multline}
where $\mu_{\mathrm{Al}}(x_{\mathrm{X}})$ and $\mu_{\mathrm{X}}(x_{\mathrm{X}})$
are the Al and X component chemical potentials in a solid solution 
of concentration $x_{\mathrm{X}}$, 
and $x^{eq}_{\mathrm{X}}$ and $x^{0}_{\mathrm{X}}$ 
the concentrations of the equilibrium and supersaturated solid solution.

LTE are more easy to handle in semi-grand-canonical ensemble 
where all quantities are written as functions 
of the effective potential $\mu=\left( \mu_{\mathrm{Al}} - \mu_{\mathrm{X}} \right)/2$.
Definition of the nucleation free energy then becomes
\begin{equation}
	\label{eq:Gnuc_GP}
	\Delta G^{\mathrm{nuc}}(\mu) = 
	\mathcal{A}(\mu^{\mathrm{eq}}) - \mathcal{A}(\mu) 
	+ \frac{1}{2}\left( \mu^{\mathrm{eq}} - \mu \right) ,
\end{equation}
where $\mu^{\mathrm{eq}}$ is the effective potential corresponding 
to equilibrium between the Al solid solution and the Al$_3$X L1$_2$ compound.
We have defined in Eq.~\ref{eq:Gnuc_GP} the solid solution semi-grand-canonical free energy
$\mathcal{A}=\left( \mu_{\mathrm{Al}} + \mu_{\mathrm{X}} \right)/2
= F(x) +\left( 1-2x \right)\mu$, $F(x)$ being the usual canonical free energy.

A LTE consists in developing the partition 
function of the system around a reference state, keeping in the series
only the excited states of lowest energies. 
Use of the linked cluster theorem \cite{DUCASTELLE,DOMB3} allows then 
to express the corresponding semi-grand canonical free energy as
\begin{equation}
	\label{eq:A_BT}
	\mathcal{A}(\mu) = \mathcal{A}^0(\mu) - kT \sum_{i,n}{g_{i,n}
	\exp{\left(-\Delta E_{i,n}(\mu)/kT\right)}},
\end{equation}
where the energy of the ground state is $\mathcal{A}^0(\mu)=U_{\mathrm{Al}} + \mu $
for the Al solid solution and $\mathcal{A}^0(\mu)=3/4\ U_{\mathrm{Al}} + 1/4\ U_{\mathrm{X}}
+ 3\oun +\mu/2$ for the Al$_3$X L1$_2$ compound.
In the sum appearing in Eq.~\ref{eq:A_BT}, the excited states
have been gathered according to their energy state $i$
and the number $n$ of lattice sites involved.
LTE parameters corresponding to the excited states with the lowest energies
are given in Tab.~\ref{tab:coef_LT}.
All excitation energies only involve a set of isolated atoms or in second nearest neighbor position
since flipping two atoms at nearest neighbor position produces an excited state
with a much higher energy.

\begin{table}[hbtp]
	\caption{Coefficients entering in the low temperature expansion
	(Eq.~\ref{eq:A_BT}). 
	The first seven excited states are considered for the solid solution 
	and the first three excited states for the Al$_3$X L1$_2$ compound.
	The effective potential is written as 
	$\mu = (U_{\mathrm{X}}-U_{\mathrm{Al}})/2 + 6\oun + \delta\mu$.}
	\label{tab:coef_LT}
	\begin{tabular}{ccccccc}
	\hline\hline
	&         & & \multicolumn{2}{c}{Solid solution}& \multicolumn{2}{c}{L1$_2$ compound} \\
	$i$ & $n$ & & $\Delta E_{i,n}(\mu)$ & $g_{i,n}$ & $\Delta E_{i,n}(\mu)$ & $g_{i,n}$ \\
	\hline
	1 & 1 &  \includegraphics[height=2.5mm]{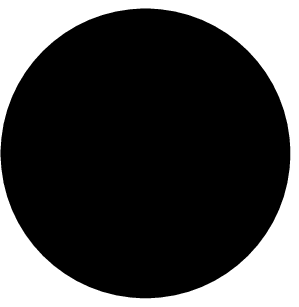}                   & $ 6\odeux-2\dmu$  & 1		& $ 6\odeux+2\dmu$  & 1/4 \\
	2 & 2 &  \includegraphics[height=2.5mm]{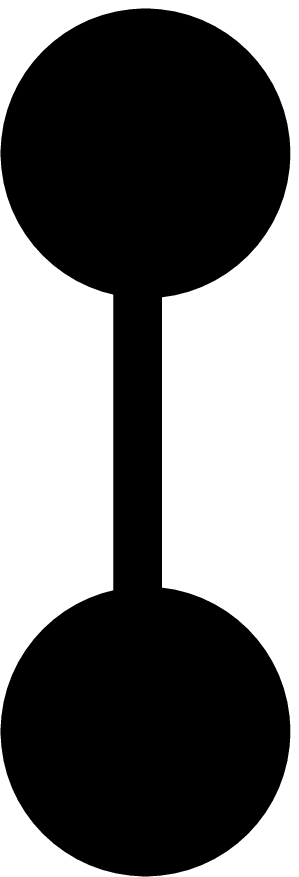}                   & $10\odeux-4\dmu$  & 3		& $10\odeux+4\dmu$  & 3/4 \\
	3 & 2 &  \includegraphics[height=2.5mm]{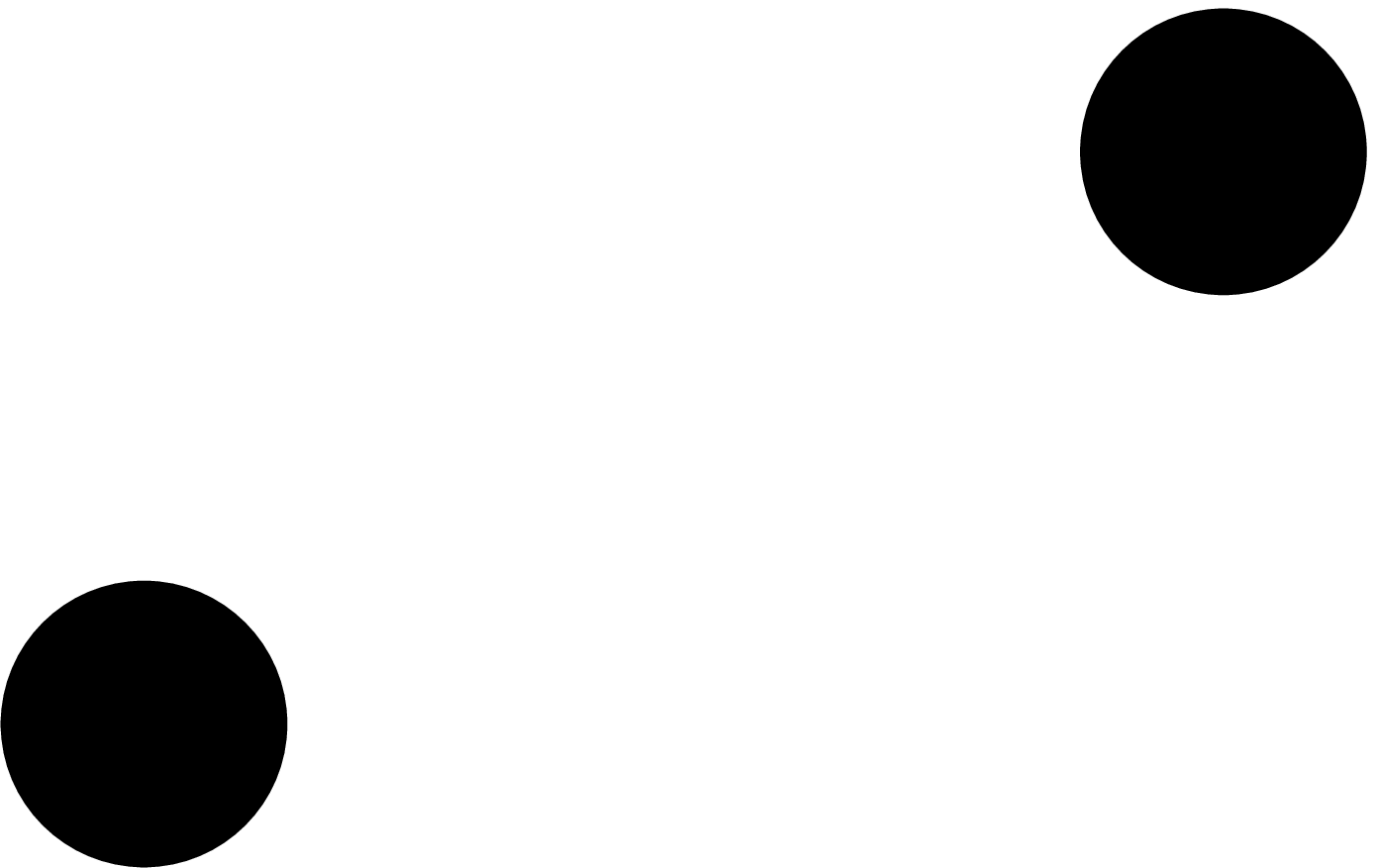}          & $12\odeux-4\dmu$  & $-19/2$	& $12\odeux+4\dmu$  & $-7/8$\\
	4 & 3 &  \includegraphics[height=2.5mm]{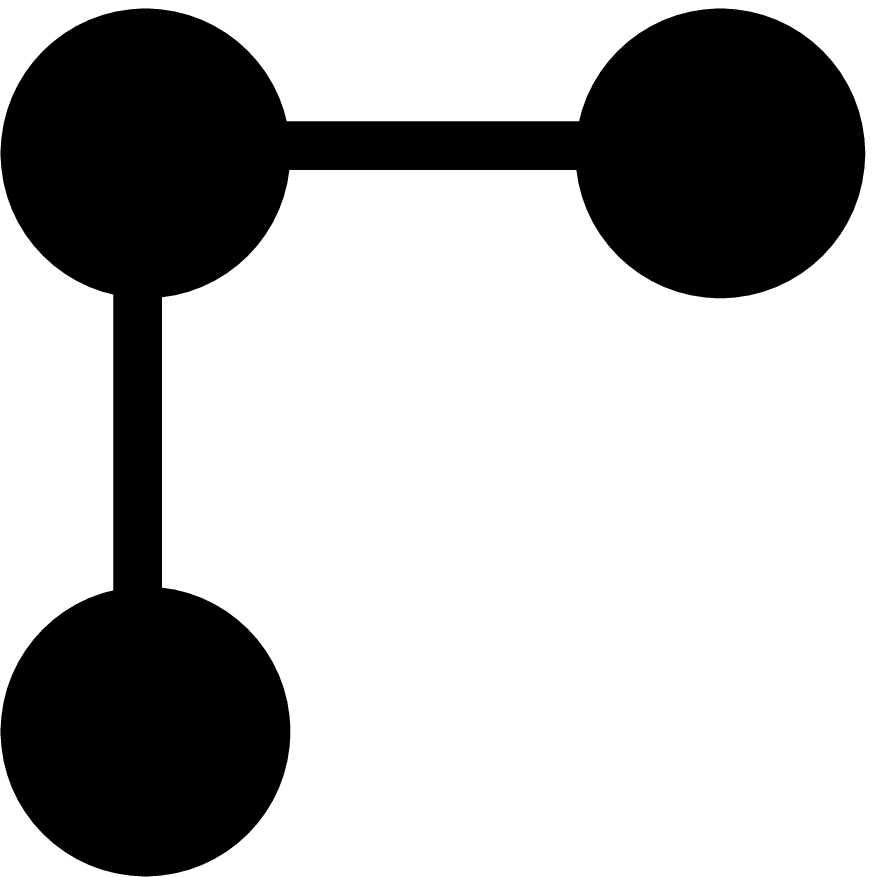}                   & $14\odeux-6\dmu$  & 15 &&\\
	5 & 4 &  \includegraphics[height=2.5mm]{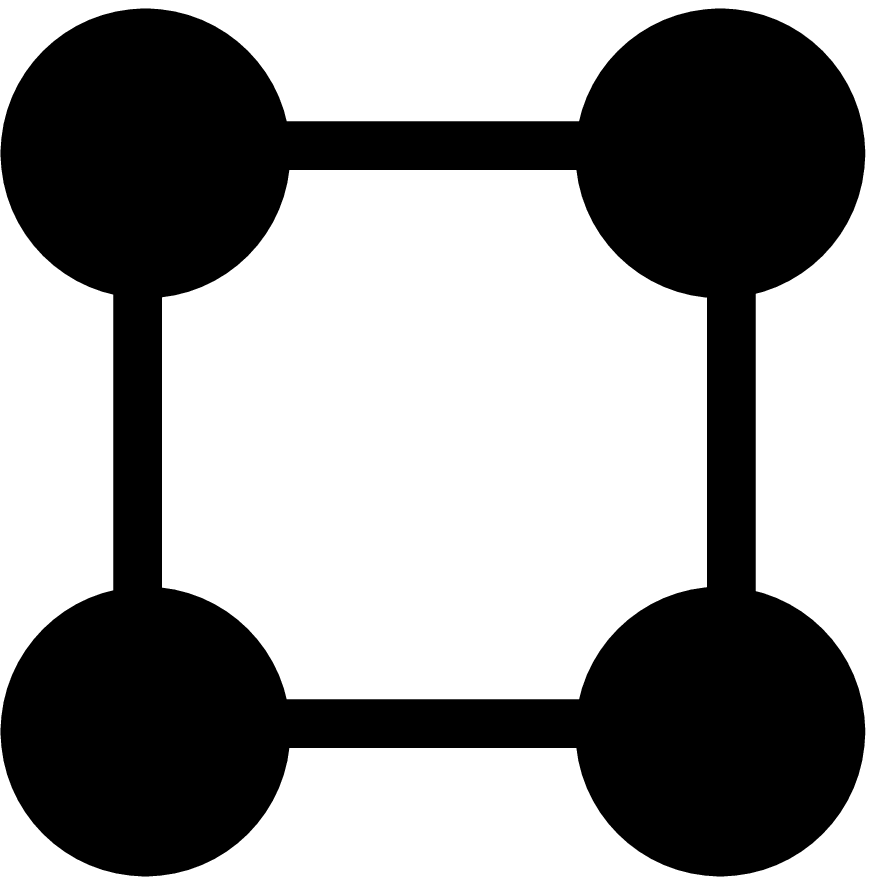}                  & $16\odeux-8\dmu$  & 3 &&\\
  	5 & 3 &  \includegraphics[height=2.5mm]{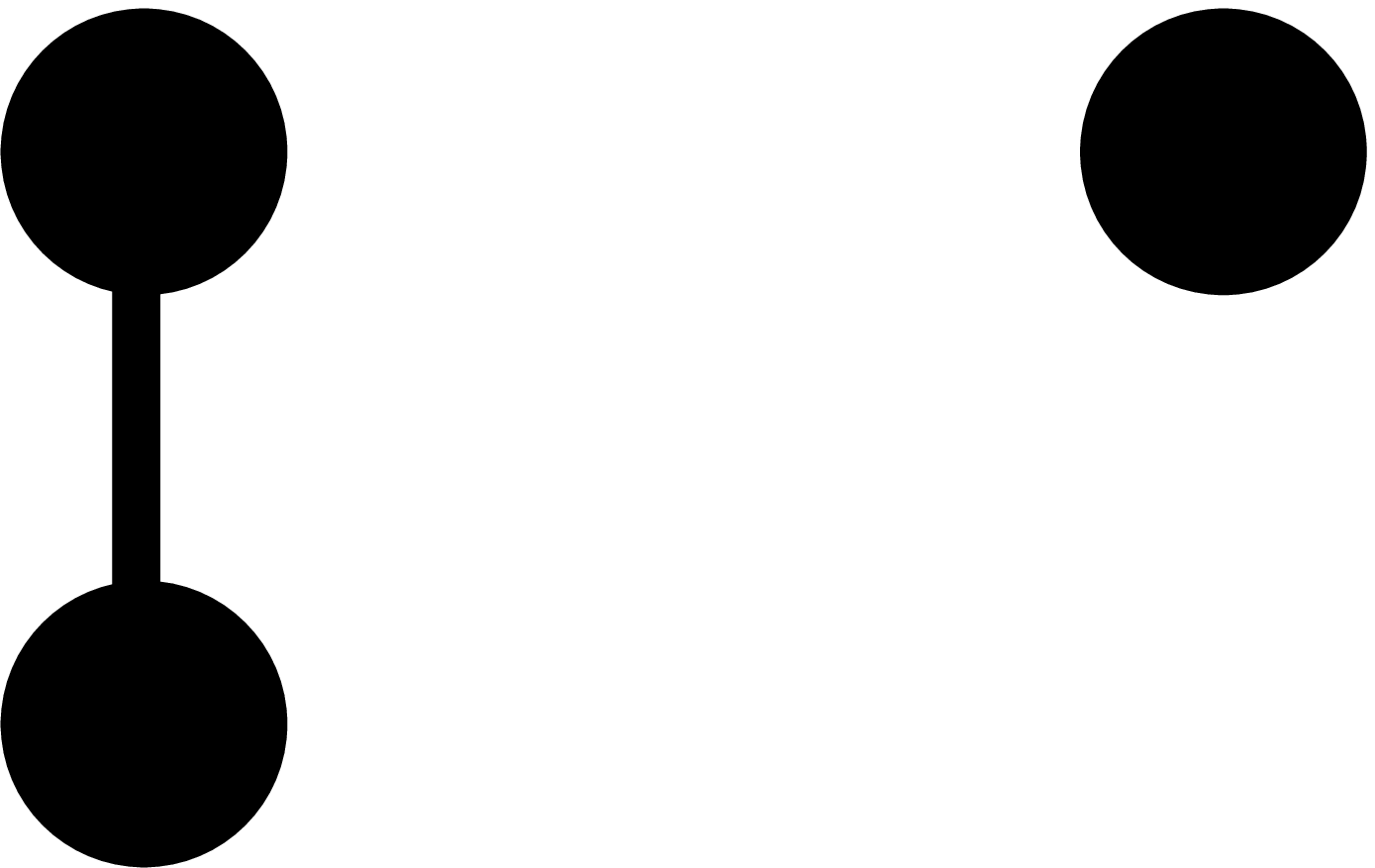}          & $16\odeux-6\dmu$  & $-96$ &&\\
	6 & 4 &  \includegraphics[height=2.5mm]{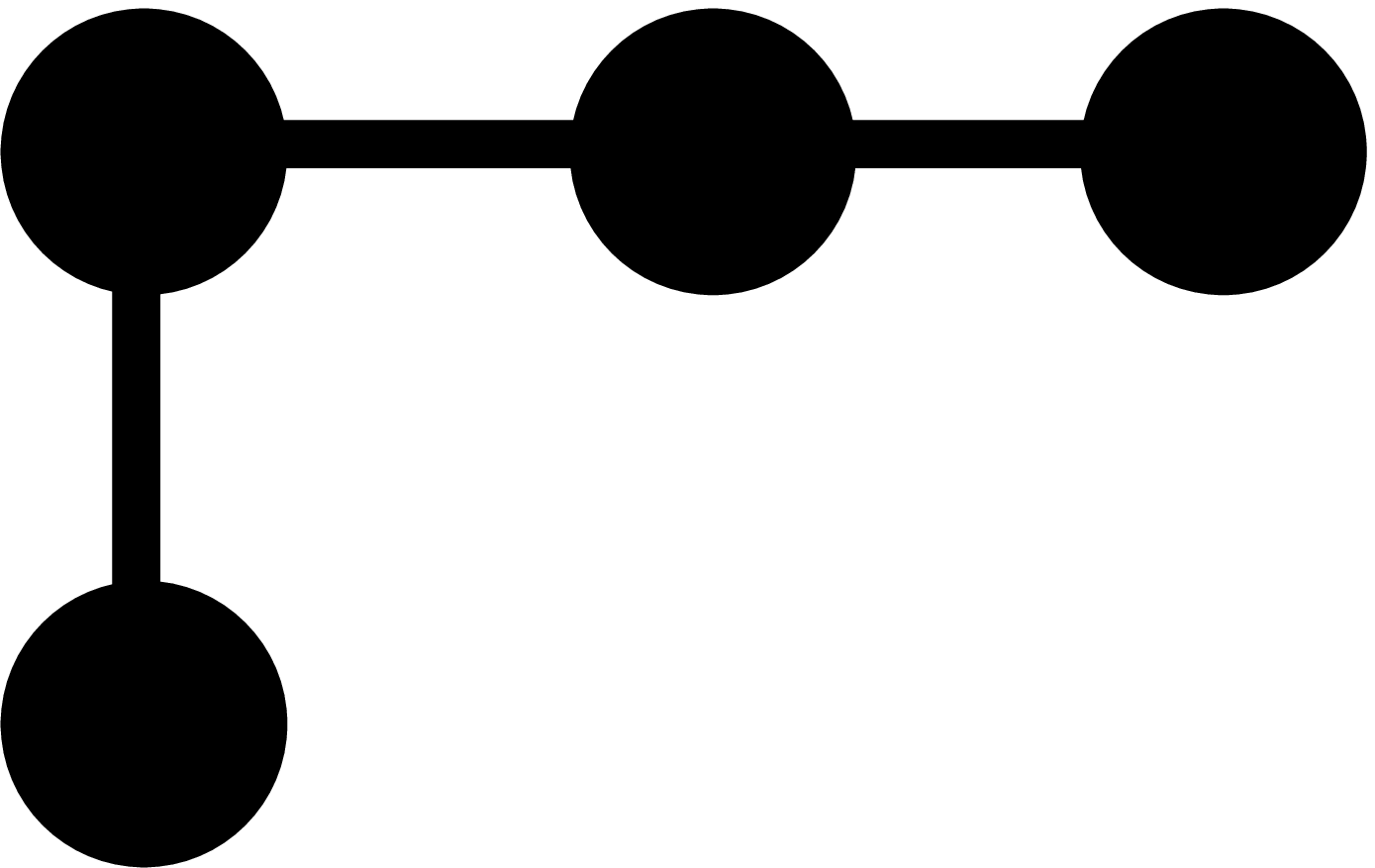}                  & $18\odeux-8\dmu$  & 83 &&\\
  	6 & 3 &  \includegraphics[height=2.5mm]{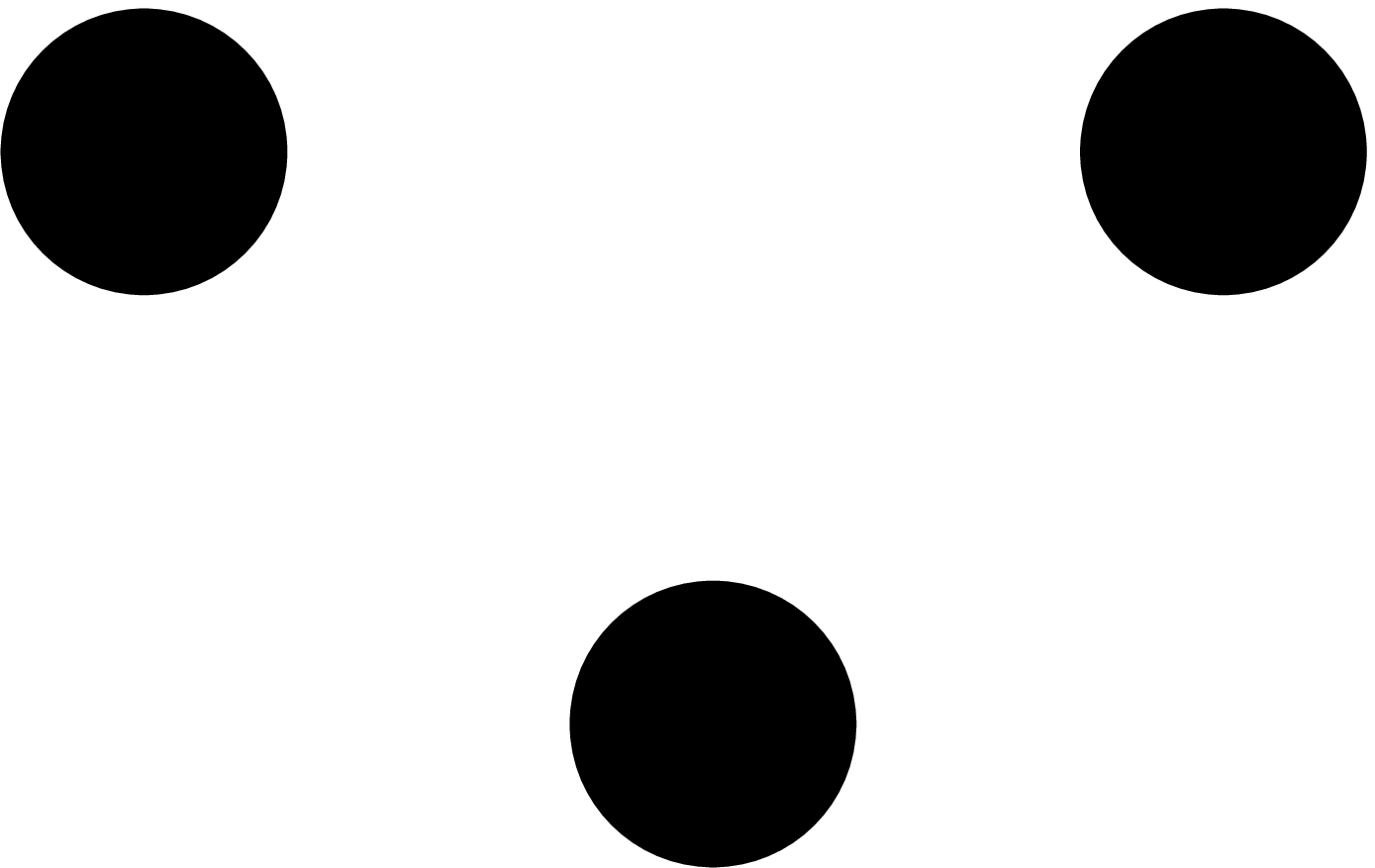} & $18\odeux-6\dmu$  & $-774$ &&\\
	7 & 5 &  \includegraphics[height=2.5mm]{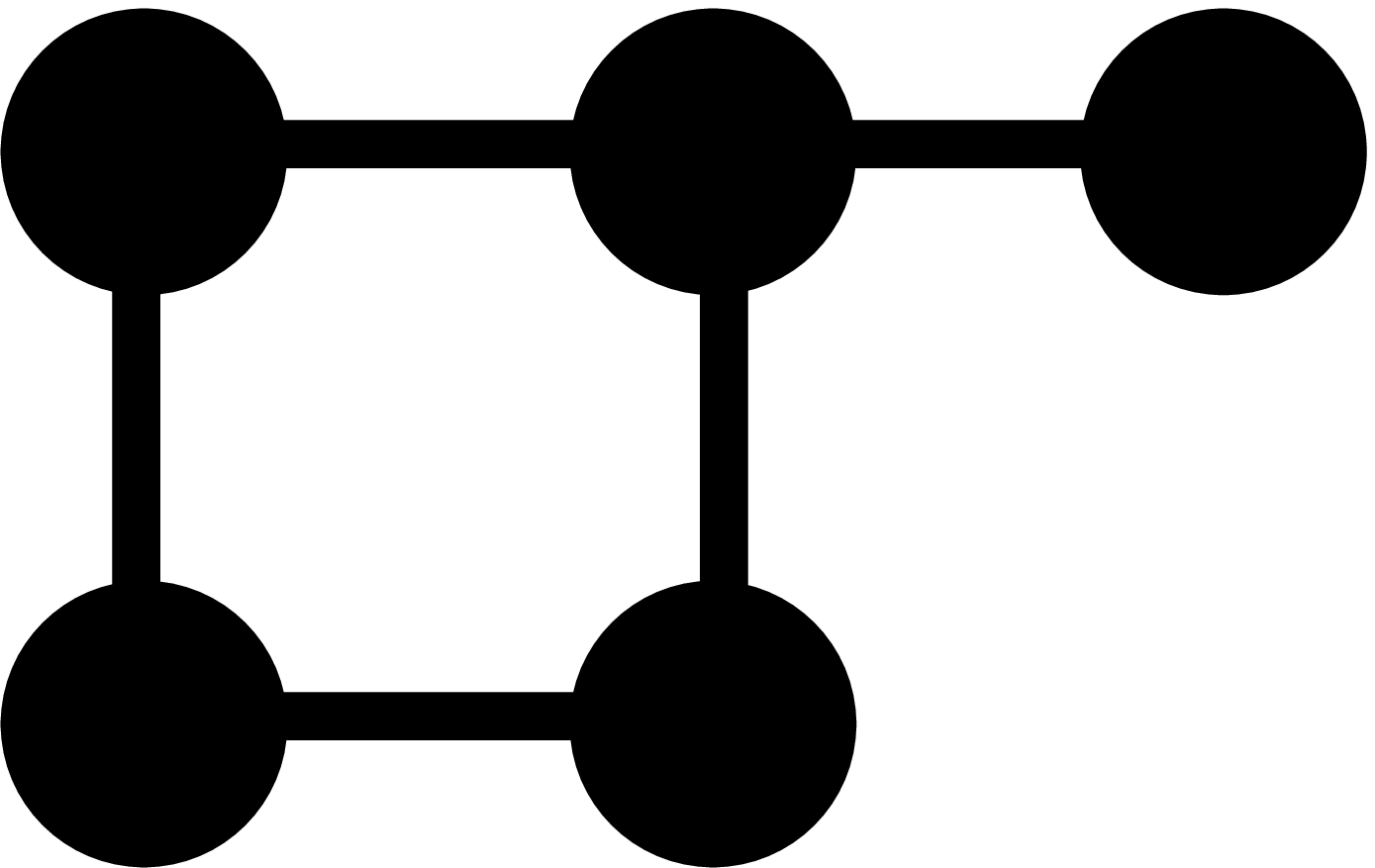}                   & $20\odeux-10\dmu$ & 48 &&\\
  	7 & 4 &  \includegraphics[height=2.5mm]{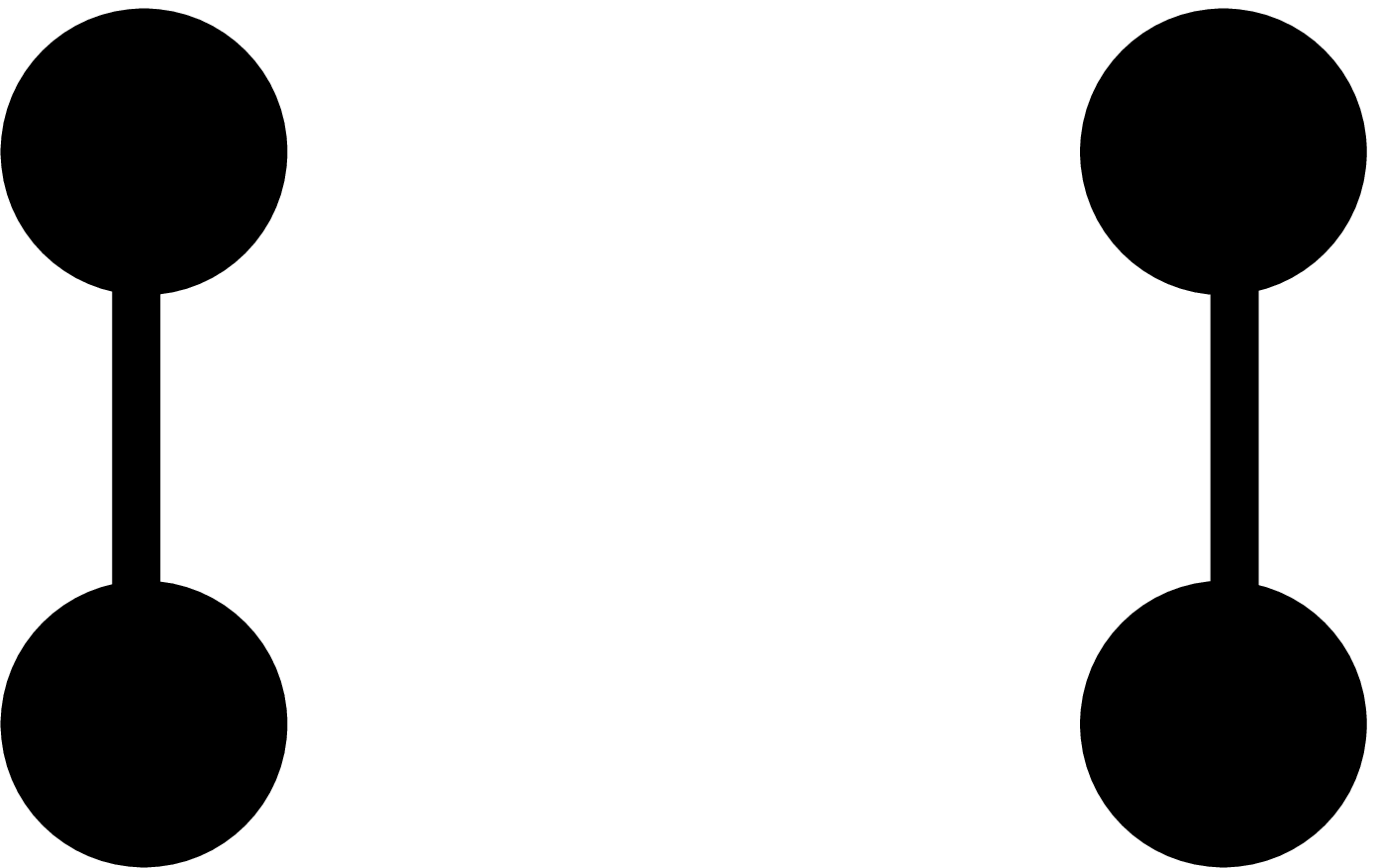} +
      		\includegraphics[height=2.5mm]{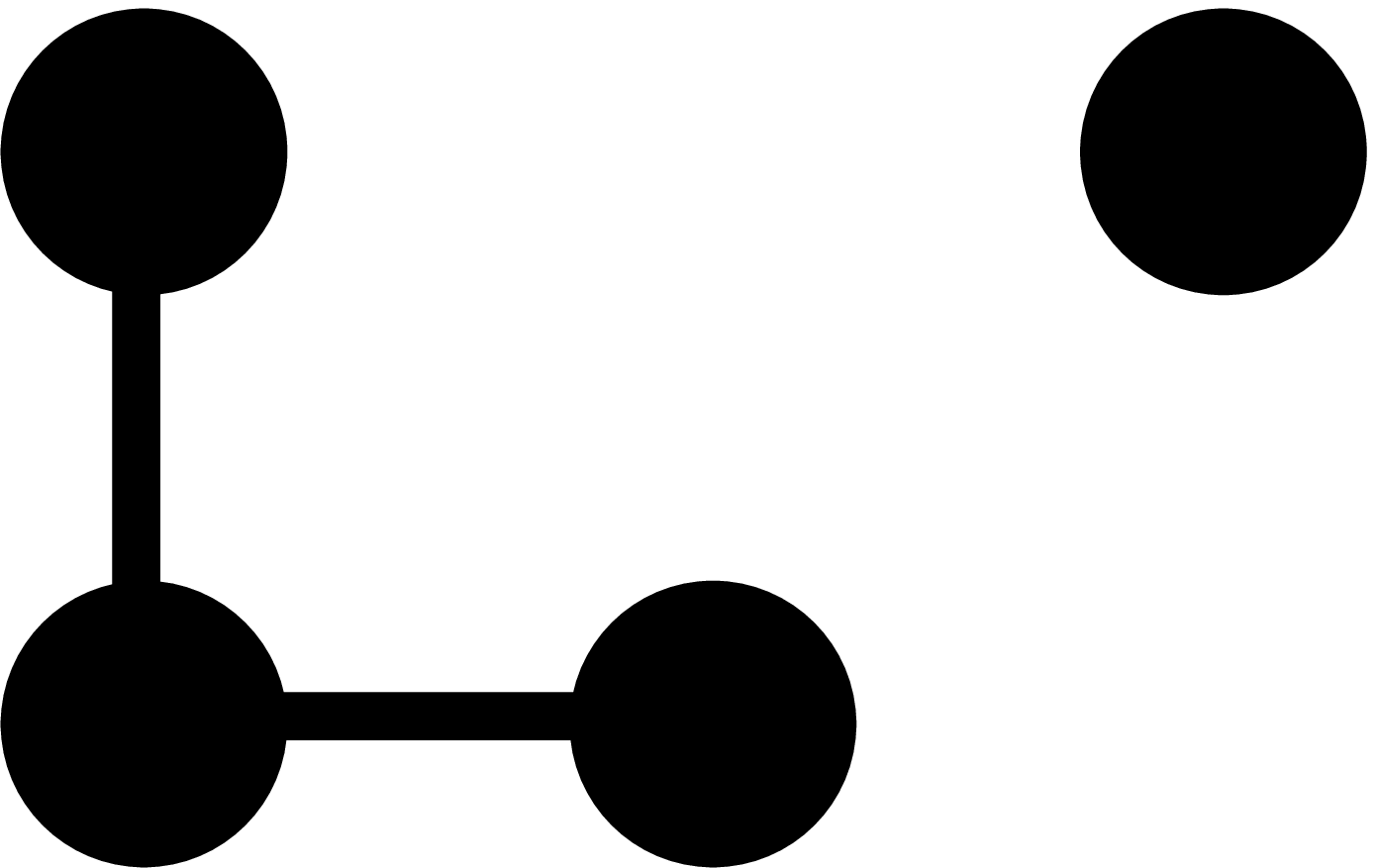}          & $20\odeux-8\dmu$  & $-1569/2$ &&\\
	\hline\hline
	\end{tabular}
\end{table}

The solute concentration in a given phase is obtained
by considering  the derivative of the corresponding semi-grand-canonical free energy.
For the solid solution, one gets
\begin{equation}
	\label{eq:conc_sol_sol_BT}
	\begin{split}
	x_{\mathrm{X}}(\mu) 
		&= \frac{1}{2}\left( 1 - \frac{\partial 	\mathcal{A}(\mu)}{\partial \mu} \right) \\
		&= \sum_{i,n}{ n g_{i,n} \exp{\left(-\Delta E_{i,n}(\mu)/kT\right)}}.
	\end{split}
\end{equation}

\begin{figure}[t]
	\begin{center}
		\includegraphics[width=\linewidth]{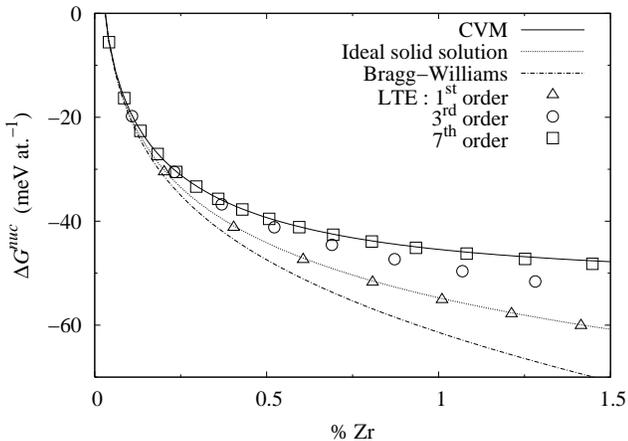}
	\end{center}
	\caption{Variation with the nominal concentration $x_{\mathrm{Zr}}^{0}$
	of the nucleation free energy $\Delta G^{\mathrm{nuc}}$ at $T=723$~K
	obtained with different thermodynamic approximations: CVM, ideal solid solution,
	Bragg-Williams and low-temperature expansions (LTE) to different orders.}
	\label{fig:Gnuc}
\end{figure}

The solid solution and the L1$_2$ compound are in equilibrium when
both phases have the same semi-grand canonical free energy.
Considering third order LTE ($i=3$), this happens for the effective potential
\begin{multline}
	\mu^{\mathrm{eq}} = (U_{\mathrm{X}}-U_{\mathrm{Al}})/2 + 6\oun \\ 
	 +  \frac{3}{2} kT \left[ \exp{(-6\odeux/kT)} 
	 + 3\exp{(-10\odeux/kT)} \right. \\
 	 \left. -\frac{13}{2}\exp{(-12\odeux/kT)} \right] ,
	\label{eq:mu_eq_BT}
\end{multline}
corresponding to the solubility
\begin{multline}
	x^{\mathrm{eq}}_{\mathrm{X}} = \exp{\left(-6\odeux/kT\right)}
	+ 6 \exp{\left(-10\odeux/kT\right)} \\
	 -16 \exp{\left(-12\odeux/kT\right)}.
	\label{eq:solubility_BT}
\end{multline}
As these expressions have to be consistent with the expansion of $\mathcal{A}$,
terms with larger exponential arguments than $-12\odeux$ are discarded.
For equilibrium phases, one does not need to go further in the expansion than the third order. 
Indeed, thermodynamic properties are already well converged 
as the solid solution and the L1$_2$ compound in equilibrium
only slightly deviate from their respective ground states.
On the other hand, Fig.~\ref{fig:Gnuc} show that an expansion beyond the third order  
of the semi-grand canonical free energy $\mathcal{A(\mu)}$ 
of the supersaturated solid solution 
and of the corresponding concentration $x_{\mathrm{X}}(\mu)$ 
improves the convergence of the nucleation free energy.
When only the first excited state is included in the expansion, LTE leads to the same value 
of the nucleation free energy as the ideal solid solution model. 
As more excited state are included in the expansion,
the value deduced from LTE converges to the one calculated with CVM
in the tetrahedron-octahedron approximation \cite{CLO04}. 
This is to contrast to the Bragg-Williams approximation which leads to a worse
prediction of the nucleation free energy than the ideal solid solution model.

So as to understand why Bragg-Williams approximation does so bad, 
it is worth going back to the canonical ensemble.
When considering only the third order LTE,
thermodynamic quantities can be expressed as functions of the solid solution
nominal concentration $\xO$.
In particular, the nucleation free energy is given by
\begin{multline}
	\Delta G^{\mathrm{nuc}}_{\mathrm{LTE}}(\xO) 
	= kT \left[ \q{(\xO)}-\q{(\xeq)} \right] \\
	 + 3kT \exp{\left(2\odeux/kT\right)}\left[ \q{(\xO)}^2-\q{(\xeq)^2} \right] \\
	-\frac{1}{4} kT \left\{ \ln{\left[\q{(\xO)}\right]}-\ln{\left[\q{(\xeq)}\right]}
	\right\} ,
	\label{eq:Gnuc_BT}
\end{multline}
where we have defined the function
\begin{equation}
	\q{(x)}=\frac{2x}
	{1+\sqrt{1+4x\left[ 6\exp{\left(2\odeux/kT\right)} - 19 \right] }}.
	\label{eq:q_function}
\end{equation}
This expression developed 
to first order in the concentrations $\xO$ and $\xeq$ leads to
\begin{multline}
	\Delta G^{\mathrm{nuc}}_{\mathrm{LTE}}(\xO) \sim
	\frac{3}{4} kT \ln{\left( \frac{1-\xeq}{1-\xO} \right)} 
	+ \frac{1}{4} kT \ln{\left( \frac{\xeq}{\xO} \right)} \\
	+ \frac{1}{4} kT \left( 1+6\me^{2\odeux/kT} \right)\left( \xO - \xeq \right).
	\label{eq:Gnuc_BT_DL}
\end{multline}
Doing the same development for the nucleation free energy calculated 
within the Bragg-Williams approximation \cite{CLO04}, we obtain
\begin{multline}
	\Delta G^{\mathrm{nuc}}_{\mathrm{BW}}(\xO) \sim 
	\frac{3}{4} kT \ln{\left( \frac{1-\xeq}{1-\xO} \right)}
	+ \frac{1}{4} kT \ln{\left( \frac{\xeq}{\xO} \right)} \\
	+ \left( 6 \oun + 3 \odeux \right)\left( \xO - \xeq \right).
	\label{eq:Gnuc_BW_DL}
\end{multline}
Comparing Eq.~\ref{eq:Gnuc_BT_DL} with Eq.~\ref{eq:Gnuc_BW_DL}, we see that 
these two thermodynamic approximations deviate from the ideal solid solution
model by a distinct linear term.
In the LTE (Eq.~\ref{eq:Gnuc_BT_DL}), the nucleation free energy is only
depending on the second nearest neighbor interaction and the coefficient
in front of the concentration difference is positive.
On the other hand, the Bragg-Williams approximation (Eq.~\ref{eq:Gnuc_BW_DL})
incorporates both first and second nearest neighbor interactions 
into a global parameter $\omega_{\mathrm{AlX}}=6\oun+3\odeux$.
This leads to a linear correction with a coefficient
which can be negative due to the oscillating nature of the interactions.
In particular, this is the case for both binary Al-Zr and Al-Sc alloys \cite{CLO04}.
Bragg-Williams approximation thus leads to a wrong correction 
of the ideal model because it does not consider properly short range order.
In the case of a L1$_2$ ordered compound precipitating from a solid solution
lying on a fcc lattice, one cannot use such an approximation to calculate 
the nucleation free energy.
On the other hand, Eq.~\ref{eq:Gnuc_BT} is a good approximation
and can be used to calculate the nucleation free energy
even when the second nearest neighbor interaction $\odeux$ is not 
known. Indeed, this parameter can be deduced from the solubility limit
$\xeq$ by inverting Eq.~\ref{eq:solubility_BT}, leading to the relation
\begin{equation}
	\odeux=-\frac{1}{6}kT\ln{\left( \xeq \right)}
	+ kT\left( {\xeq}^{2/3} - \frac{8}{3}{\xeq} \right).
	\label{eq:omega2_BT}
\end{equation}
This relation combined with Eq.~\ref{eq:Gnuc_BT} provides a powerful
way for calculating the nucleation free energy from the solid solubility.

LTE can be used too to calculate the plane interface free energy $\sigma_{100}$
corresponding to a [100] direction. 
Due to the inhomogeneity perpendicular to the interface, 
the main contribution arises from broken bonds 
and excited states, whose energies are lower than in bulk phases, 
only bring a small correction. 
One thus does not need to go further than the second order in the expansion. 
At 0~K, the isotropic interface free energy $\bar{\sigma}$
is obtained by multiplying $\sigma_{100}$ with the geometric factor 
$\left( 6/\pi \right)^{1/3}$ corresponding to a perfect [100] facetting of the precipitates. 
For low temperatures, this is a good approximation to assume 
that the same linear relation holds between both quantities \cite{CLO04}. 
The isotropic interface free energy given by LTE is then
\begin{multline}
	\label{eq:sigma}
	a^2\bar{\sigma} = \left( 6/\pi \right)^{1/3}\left[ 
	\odeux - 2 kT \exp{(-4\odeux/kT)} \right. \\
	\left. - kT  \exp{(-6\odeux/kT)} \right],
\end{multline}
where $a$ is the fcc lattice parameter.

\begin{figure}[t]
	\begin{center}
		\includegraphics[width=\linewidth]{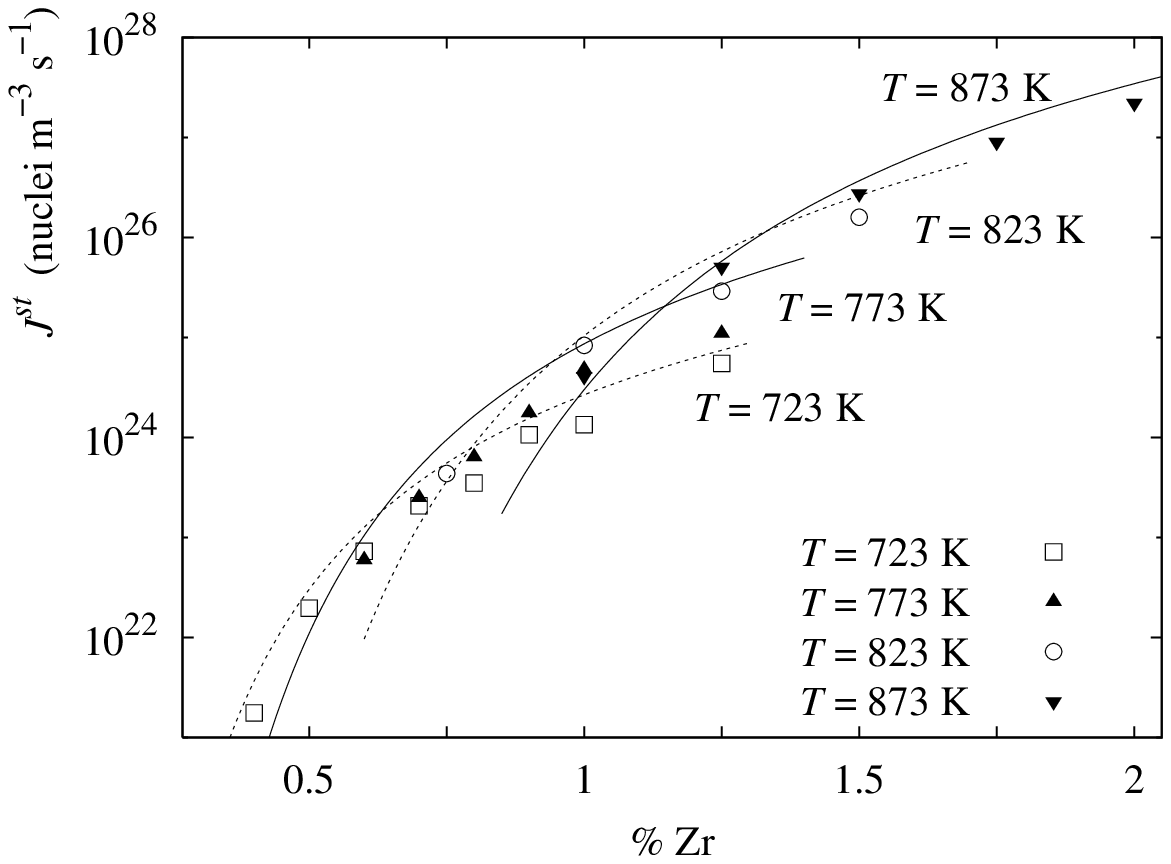}
		\includegraphics[width=\linewidth]{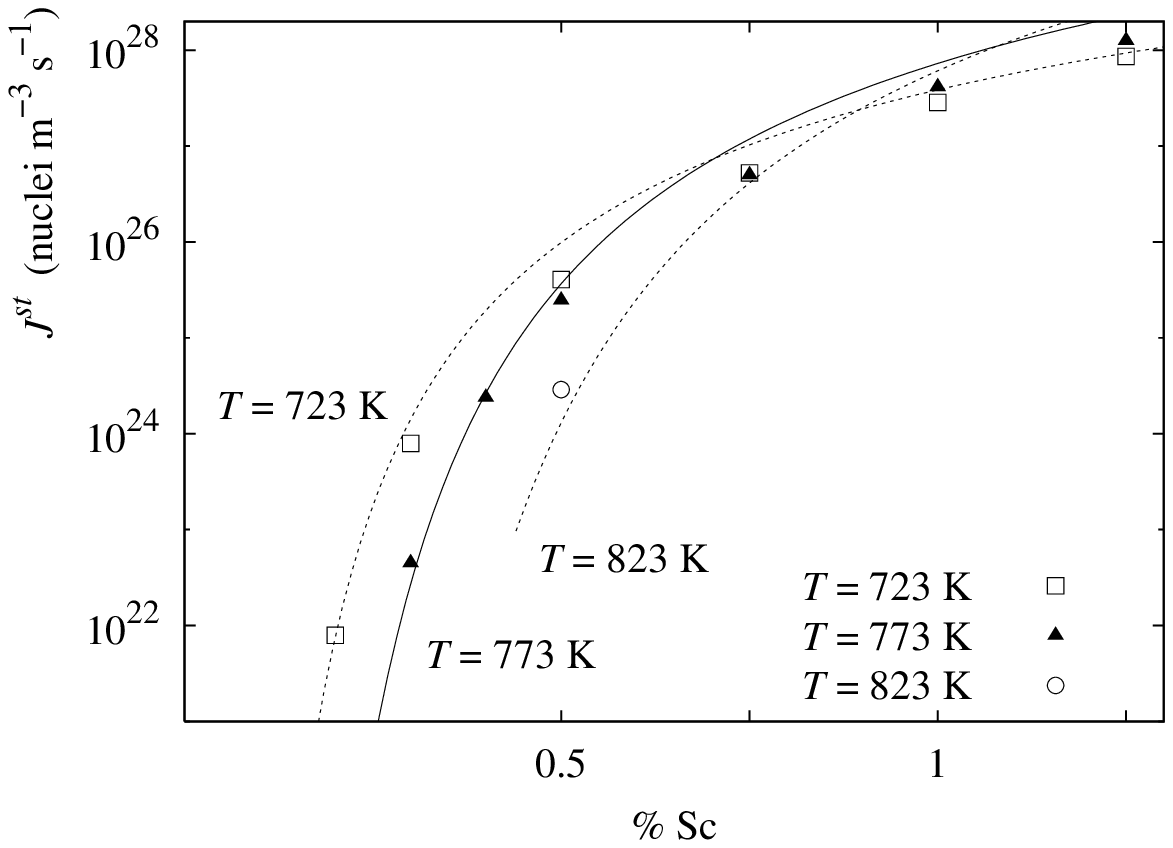}
	\end{center}
	\caption{Variation with nominal concentration and temperature 
	of the steady-state nucleation rate $J^{\mathrm{st}}$ for Al$_3$Zr (top)
	and Al$_3$Sc (bottom) precipitations. 
	Symbols correspond to kinetic Monte Carlo simulations and lines 
	to classical nucleation theory}
	\label{fig:Jst}
\end{figure}

LTE thus allow to calculate all CNT input parameters
from the knowledge of the solubility limit.
The nucleation rate is then obtained from the equation 
\begin{multline}
	J^{\mathrm{st}}(x_{\mathrm{X}}^0) 
	= -16 N_s \frac{\Delta G^{\mathrm{nuc}}(x_{\mathrm{X}}^0)}{\sqrt{kTa^2\bar{\sigma}}}
	\frac{D_{\mathrm{X}}}{a^2} x^0_{\mathrm{X}} \\
	\exp{\left( -\frac{\pi}{3}\frac{(a^2\bar{\sigma})^3}
		{kT[\Delta G^{\mathrm{nuc}}(x_{\mathrm{X}}^0)]^2} \right)} ,
	\label{eq:Jst}
\end{multline}
where $D_{\mathrm{X}}$ is the X impurity diffusion coefficient in Al.
We thus obtain a fully analytical expression of the nucleation rate. 
Using the same experimental data, \ie solubility limits and diffusion coefficients,
as the ones used to fit the atomic diffusion model of kinetic Monte Carlo simulations,
we can compare CNT predictions with nucleation rate observed in simulations \cite{CLO04}.
A good agreement is obtained both for Al-Zr and Al-Sc binary alloys (Fig.~\ref{fig:Jst}).

The combination of LTE with CNT thus allows to build a quantitative modeling
of nucleation relying on a very limited number of material parameters. 
Such a model can be directly applied to aluminum alloys
where a L1$_2$ compound precipitates from a supersaturated solid solution
as this is the case with Zr, Sc or other rare earth elements like Er, Tm, Yb and Lu \cite{KNI06}.
Li too precipitates in aluminum with a L1$_2$ structure, 
but this system requires another statistical approximation than LTE.
Indeed, this approach based on LTE,
requires that the precipitating phase only slightly deviates from its perfect stoichiometry 
and that the solute solubility remains low.
Provided these conditions are fulfilled, it could be applied to alloys other than aluminum alloys.
More generally, LTE demonstrate that the oscillating nature 
of the interactions in an alloy with an ordering tendency 
has to be taken into account by the CNT and requires a better
statistical description than the Bragg-Williams approximation
which treats all interactions on the same footing.

	The authors would like to thank Y. Le Bouar and A. Finel
	for helpful discussions on LTE, and B. Legrand, F. Soisson
	and G. Martin for their invaluable help.

\end{document}